\documentclass[aps,prb,preprint,grouppedaddress]{revtex4}

\usepackage{graphicx}
\usepackage{dcolumn}
\usepackage{amsmath}
\usepackage{multirow}
\usepackage{epsfig}
\usepackage{amssymb}
\usepackage{color}

\begin{document}

\title{Strong enhancement of magnetic order from bulk to stretched monolayer FeSe
as Hund's metals}

\author{Chang-Youn Moon}
\email{cymoon@kriss.re.kr}
\affiliation{Advanced Instrumentation Institute, Korea Research Institute of Standards and Science, 
Yuseong, Daejeon 305-340, Republic of Korea. email: cymoon@kriss.re.kr}

\date{\today}

\begin{abstract}
Despite of the importance of magnetism in possible relation to other key properties 
in iron-based superconductors, its understanding is still far from complete especially for FeSe systems.
On one hand, the origin of the absence of magnetic orders in bulk FeSe is yet to be clarified. 
On the other hand, it is still not clear how close monolayer FeSe on SrTiO$_3$,
with the highest transition temperature among iron-based superconductors, 
is to a magnetic instability. Here we investigate
magnetic properties of bulk and monolayer FeSe using dynamical mean-field
theory combined with density-functional theory. We find that suppressed magnetic 
order in bulk FeSe is associated with the reduction of inter-orbital charge fluctuations, 
an effect of Hund's coupling, enhanced by a larger crystal field splitting. 
Meanwhile, spatial isolation of Fe atoms 
in expanded monolayer FeSe leads into a strong magnetic order, which is completely
destroyed by a small electron doping. Our work provides a comprehensive understanding 
of the magnetic order in iron-based superconductors and other general multi-orbital correlated
systems as Hund's metals.
\end{abstract}


\maketitle

\begin{flushleft}
{\large {\bf Introduction}}
\end{flushleft}
Magnetism is one of universal features found in iron-based superconductors (IBS)
as superconductivity generally appears in the vicinity of antiferromagnetic (AFM) 
phase with a specific stripe-type ordering pattern, from which electron
pairing mechanisms of the magnetic origin were introduced \cite{Mazin2,Zhang,Lee2,Ji,Dai,Yin2014,Allan}.
Furthermore, nematicity (spontaneous breaking of four-fold rotational symmetry of tetragonal
phase), magnetism, and superconductivity 
in IBS are thought to be closely related \cite{Fernandes,Yamakawa,Matsuura,JLi}.
In this context, understanding magnetism can be a starting point to unravel the complex inter-dependence
of these properties.
In terms of magnetism, FeSe holds a unique position among general IBS
as bulk FeSe has no magnetic ordered phase unlike most of other 
materials \cite{Hsu,McQueen09prb,Baek}, whose underlying mechanism is still not well understood. 
FeSe is also of great interest due to the highest superconducting transition
temperature among IBS when its monolayer (ML) is on SrTiO$_3$ substrate \cite{QYWang,SHe,STan,JFGe}. 
Whether or not ML FeSe/SrTiO$_3$ is close to a magnetic instability is therefore an intriguing
question. 

Meanwhile, there is a general consensus that the electron correlation should be taken into account
to properly understand material properties of this system \cite{Haule1,HauleNJP,MYi1,MYi2}. Since it is a 
multi-orbital system 
in which all five $d$ orbital bands are crossing or near the Fermi energy ($E_F$), Hund's coupling 
$J_H$ is an indispensable part of interactions as well as the intra-orbital Coulomb repulsion $U$, 
and IBS in the correlated metallic state are often described as Hund's metals \cite{HauleNJP,YinNMAT,Fanfarillo1,Medici}.
In this material state, reduced inter-orbital
Coulomb repulsion $U' = U - 2J_H$ \cite{Kanamori}
and the tendency to promote parallel spin alignment cooperatively 
decouple the five $d$ orbital components, which is signaled by the suppression of inter-orbital charge 
fluctuations. Consequently, coherent and 
incoherent states can coexist and some orbitals are close to Mott transition while the others
are still itinerant. Since this orbital selectivity is known to be enhanced in FeSe \cite{MYi1,MYi2,Aichhorn,YinNMAT,Kostin},
its magnetic properties would be better understood in the context of Hund's metal physics.

In this work, a systematic comparative study on the magnetic properties of FeSe in different forms
and a reference IBS, LaFeAsO, is performed using a density-functional
theory plus dynamical mean-field theory (DFT+DMFT). 
It is found that the inter-orbital charge fluctuations are greatly reduced between 
$e_g$ and $t_{2g}$ orbitals for bulk FeSe due to its large
crystal field splitting and the resultant strong orbital decoupling induced by the 
Hund's coupling. Consequently the total charge fluctuation 
are enhanced leading to a largely reduced ordered magnetic moment 
compared with LaFeAsO, consistently with the absence of magnetic order in bulk FeSe 
in experiments. In contrast, increased fluctuating magnetic moment
and suppressed total charge fluctuation due to the increased inter-atomic distance and
the reduced dimentionality result in a large ordered 
magnetic moment in expanded ML FeSe with the lattice 
constant of that on SrTiO$_3$. Thus, the stark contrast of the magnetic order
between bulk and ML FeSe is explained in terms of Hund' metal properties within an unified framework.
Small electron doping is found to effectively destroy the magnetic order 
in this system, implying that the superconductivity in ML FeSe/SrTiO$_3$ is in the vicinity of magnetic 
order. 

\begin{flushleft}
{\large {\bf Results}}
\end{flushleft}
{\bf Magnetic susceptibility.} Three different materials are considered in this work; namely, 
LaFeAsO as an archetypal IBS,
bulk FeSe, and freestanding ML FeSe tensile-strained to the lattice constant of
ML FeSe/SrTiO$_3$, 3.90 \AA~\cite{STan}.
A recent DFT+DMFT study demonstrated that the main effect of defect-free SrTiO$_3$ 
substrate on the electronic structure of ML FeSe is to increase 
the Se-Fe-Se angle through increasing the lattice
constant of ML FeSe \cite{Mandal}, and an earlier DFT study suggested a similar conclusion \cite{HCao1}.
Therefore, strained freestanding ML FeSe is expected to
capture most of the essential features of magnetic properties of that on SrTiO$_3$ as well.
Electron doping, another important possible substrate effect, will be also discussed in the later part
of this work. 

Figure 1 displays the imaginary part of the magnetic susceptibility, 
$\chi_m''$, as a function of momentum and frequency, for the three materials in the paramagnetic (PM) phase.
The magnetic susceptibility is estimated within DFT+DMFT method from the Bethe-Salpeter equation, using
fully momentum and frequency dependent interacting DFT+DMFT one-particle lattice Green's function and local
two-particle vertex function obtained from the DMFT impurity solver \cite{Hyowon}.
$\chi_m''$ for LaFeAsO
exhibits a typical spin excitation spectrum for IBS, with largest weights 
at $q=(1,0)$ near zero frequency indicating the magnetic instability for the stripe-type AFM order
and also with high energy excitations near $q=(1,1)$, as can be seen in previous
similar calculations \cite{Hyowon,Yin2014,Chenglin,Moon2016}. Meanwhile, low energy spin fluctuations
are much suppressed for bulk FeSe indicating the weakened tendency for the magnetic order
in accordance with its absence in experiments. Considerable amount of spectral weights 
near zero energy are relocated to near 100 meV, implying that some higher frequency processes are involved in 
the magnetic order suppression. Finally, ML FeSe exhibits the overall increase of spectral weights as well
as the recovered dominance of low energy fluctuations over high energy ones compared with bulk FeSe,
indicating a much stronger tendency for the magnetic order. In this case, however, strongest low
energy excitations are not at $q=(1,0)$, but slightly shifted from it toward $q=(1,1)$ 
suggesting an incommensurate
magnetic order. Besides the spin fluctuation, as the orbital degree of freedom is considered another candidate 
to drive the nematic order and/or the superconductivity in these materials, we also estimate the orbital
susceptibility (see Supplementary Figure 1 and Note 1). Only very weak low energy excitations are 
found for all the three materials, indicating
that DFT+DMFT method does not support the existence of orbital orders in these materials.

\begin{flushleft}
\end{flushleft}
{\bf Trends in local quantities.} We perform a systematic analysis for the trend of local correlations 
to understand the properties found in the susceptibility
results. Ordered magnetic moment $\langle S_z \rangle$ on a Fe atom is estimated in
the stripe-type AFM phase and found to vary from 0.70 to 0.43 and
1.00 $\mu_B$ for LaFeAsO, bulk and ML FeSe, respectively. 
We can see that the magnetic order
is suppressed and then greatly enhanced for bulk and ML FeSe compared with LaFeAsO as predicted
by magnetic susceptibility results in the PM phase in Fig.~1, and also in qualitative
agreement with the experimental observation of no magnetic order for bulk FeSe. 
Usually magnetic order is strong in materials with strong electron correlation,
and greatly reduced ordered moment 
of bulk FeSe is rather puzzling since it is considered to be more correlated than LaFeAsO. 
Indeed, mass enhancement factor, $1/Z = 1-\frac{\partial\Sigma(\omega)}{\partial\omega}\rvert_{\omega=0}$,
is found to increases considerably for $t_{2g}$ orbitals, especially $d_{xy}$ as shown in Fig.~2a. 
Although $e_g$ orbitals become less correlated from LaFeAsO to bulk FeSe, the fluctuating magnetic moment 
($\langle S^2 \rangle^{1/2}$)
which reflects the overall correlation strength, slightly increases in Fig.~2b suggesting that 
the suppressed magnetic order in bulk FeSe cannot be understood by the overall correlation strength
of the material. 
Meanwhile, mass enhancement increases for all the orbitals for ML FeSe in Fig.~2a
along with the fluctuating moment in Fig.~2b defining this material most correlated among the three. 

Using the same $U's$ and $J's$ for all the materials in the present study (see Supplementary Note 2), 
the variation of correlation strength can be
attributed mainly to that of the inter-atomic distance and orbital occupations. In spite of the large
reduction of the Fe-anion distance from 2.42 to 2.39 \AA~for LaFeAsO and bulk FeSe, respectively,
$t_{2g}$ orbitals become much more correlated while $e_g$ orbitals exhibit the opposite behavior
to produce a large orbital differentiation in bulk FeSe. As can be seen in Fig.~2c, it
results from the large difference of occupation numbers between $t_{2g}$ and $e_g$ 
orbitals which are essentially decoupled in a Hund's metal \cite{HauleNJP,Fanfarillo1,Medici}, 
indicating a large crystal field splitting in bulk FeSe. We estimate that all five Fe-$d$ orbital  
levels lie within the range of 0.25 eV for LaFeAsO while the range increases to 0.48 eV for 
bulk FeSe indeed confirming the enhanced crystal field splitting in bulk FeSe.          
Noteworthy is that even in bulk FeSe the crystal field splitting is smaller than $J$ value
adopted in this work, 0.8 eV, so that the Hund's coupling still plays a major role in the local correlation
over all five $d$ orbitals in this material. The overall increase of mass enhancement of ML FeSe can then be
related to the elongation of Fe-anion bond to 2.40 \AA~due to the applied strain, considering 
that its orbital occupations do not change much from those of bulk FeSe. 
Also, the kinetic energy reduction in a two-dimensional system is expected to further contribute 
to the stronger overall correlation in ML FeSe, especially for $d_{xz/yz}$ and $d_{z^2}$ orbitals.

In a Hund's metal, the local charge fluctuation $\langle n^2 \rangle - \langle n \rangle^2$ where $n$ is the
local density operator on an atom, which quantifies
the charge delocalization, can be sizable even in the strongly correlated case because the electron
correlation comes from the dominance of high-spin states in the local subspace while electrons can hop
through unoccupied orbitals \cite{Fanfarillo1}. Hund's coupling promotes a 
fluctuating moment while this active charge fluctuation hinders its static order leading to the much reduced magnitude
of ordered moment compared with the fluctuating moment \cite{HauleNJP,YinNMAT,Chenglin}. 
Figure 2b indeed shows the correlation between the charge fluctuation and ordered moment,
where the enhanced charge fluctuation for bulk FeSe accounts for its suppressed ordered moment
of 0.43 $\mu_B$ compared with 0.70 $\mu_B$ of LaFeAsO while the suppressed charge fluctuation
coincides with the enhanced ordered moment of 1.00 $\mu_B$ for ML FeSe.

To understand the variation of charge fluctuation over materials, orbital-resolved charge
fluctuations defined as
\begin{equation}
\langle n_{\alpha}n_{\beta}  \rangle - \langle n_{\alpha} \rangle  
\langle n_{\beta} \rangle
\end{equation},
where $\alpha$ and $\beta$ are orbital
indexes, are estimated and listed in Table 1.
Diagonal elements represent intra-orbital charge fluctuations and are roughly correlated with respective
orbital occupations where orbitals close to the half-integer filling 1.5 have higher charge fluctuations.
Meanwhile, off-diagonal elements
correspond to inter-orbital charge fluctuations and have negative values, due to the inter-orbital
Coulomb repulsion $U'$. Their small (absolute) values are the signature of the orbital decoupling
which characterizes Hund's metals, and can contribute to increase the total charge fluctuation of an atom.
Larger overlap between $e_g$ and $t_{2g}$ orbitals enhances $U'$, and
hence inter-orbital charge fluctuations are dominant between them.
From LaFeAsO to bulk FeSe, intra-orbital charge fluctuation slightly increases or remain almost the same 
for $e_g$ orbitals, while it is considerably suppressed for $t_{2g}$ orbitals (by 18~\% for $d_{xy}$) so 
that there are large differences between $e_g$ and $t_{2g}$ orbitals, following the trend of orbital occupations 
shown in Fig.~2c. In contrast, inter-orbital components greatly increase (decreased absolute values),
especially between $d_{x^2-y^2}$ and $d_{xy}$ orbitals by over 50~\%, which overcomes the overall
reduction of intra-orbital components and produce the net increase of total charge fluctuation
as displayed in Fig.~2b. Hund's coupling keeps the the magnitude of the local spin on an iron atom (i.e., $S^2$) 
finite in both materials 
as indicated by their similar fluctuating moments in Fig.~2b and orbital-resolved
spin fluctuations (see Supplementary Table 1 and Note 3). Meanwhile, the charge fluctuation enhances the chance of spin flip processes of this local spin
as a whole, not losing Hund's coupling energy, to result in the contrasting 
ordered moments between LaFeAsO and bulk FeSe as shown in Fig.~2d where every orbital component of the ordered moment 
is reduced for the latter compared with the former. As mentioned earlier, the enhanced spin flip processes
reducing the ordered moment in bulk FeSe can be associated with the 100 meV spin excitations in Fig.~1.

The pronounced suppression of inter-orbital charge fluctuations in bulk FeSe can be attributed to 
the large difference 
of its intra-orbital components between $e_g$ and $t_{2g}$ orbitals shown in Table 1,
as the inter-orbital fluctuation is expected to be suppressed between orbitals which fluctuate
incoherently to each other with very different rates. Since the difference in intra-orbital 
charge fluctuations among orbitals can be mainly accounted for by that in orbital occupations as mentioned
above, their larger difference in bulk FeSe is the direct consequence of the larger crystal 
field splitting. In short, the suppressed magnetic order in bulk FeSe compared with LaFeAsO is 
a result of its relatively large crystal field splitting which produces a strong orbital selectivity
by the action of Hund's coupling (see Supplementary Note 4).
Meanwhile, orbital occupations do not change much from bulk to ML FeSe in Fig.~2c and
hence neither do inter-orbital charge fluctuations and other components in Table 1. The
decreased total charge fluctuation of ML FeSe is a cooperative result from all of the components
with small and even contributions, without a single dominant one. Together with the increased
fluctuating moment as shown in Fig.~2b, the suppressed charge fluctuation leads to a strong
magnetic order of 1 $\mu_B$ and can be considered as a natural consequence of localized orbitals
with increased inter-atomic distances compared with bulk FeSe.

\begin{flushleft}
\end{flushleft}
{\bf Effects of doping on ML FeSe.} The stabilization of the AFM phase in ML FeSe on SrTiO$_3$ 
has been also predicted by previous DFT 
calculations \cite{KLiu,HCao1,HCao2}, however, with large ordered moments of over 2 $\mu_B$ which are
likely overestimated as is a well-known general property of DFT on IBS.
A recent experimental work indeed confirmed an AFM order in this system using magnetic exchange
bias effect measurement \cite{YZhou}, though neither the ordering vector nor the ordered moment 
could be determined. It is also found that the magnetic order disappears for the electron doped
sample where superconductivity can arise.
To investigate the effect of doping on the electronic and magnetic properties of ML FeSe, 
0.12 $e^-$/Fe is added as determined on the superconducting sample by an earlier experimental 
study \cite{STan}. 
Fig.~3a and d show the spectral function A($k,\omega$) with orbital characters
in the BZ of one-Fe-atom
unitcell for undoped and electron doped ML FeSe systems, respectively. Two hole bands around $\Gamma$
and electron bands around $X$ are mainly of $d_{xz/yz}$ character, while another hole band at
$M$ is from $d_{xy}$ orbital. Meanwhile, $e_g$ orbital components are located relatively farther
from the $E_F$. Although some spectral weights are above the $E_F$ for the hole
band at $M$ due to its incoherence, its real eigenvalues which determine the peak positions of
A($k,\omega$) are actually below the $E_F$, so in the FS plot in Fig.~3b and e, 
no hole FS is shown around $M$. Around $\Gamma$, on the other hand, two small hole FS
exist for undoped ML FeSe, while they sink below the $E_F$ for the doped case. Consequently,
no hole surface is present for the doped ML FeSe, in agreement with experimental observations on
ML FeSe/SrTiO$_3$ system \cite{DLiu,SHe,STan,MYi2} as well as previous DFT+DMFT calculations
\cite{Nekrasov1,Nekrasov2}. In Fig.~3c and f, $\chi_m''(q,\omega=5$ meV) in the PM
phase is displayed to figure out how the
static magnetic order evolves with doping. In the undoped case, static order is predicted 
slightly off the stripe-type AFM ordering vector as is already seen in Fig.~1. Despite of
significant renormalization of the non-interacting susceptibility $\chi_0$ by the local 
two-particle vertex to form the fully interacting $\chi$ \cite{Chenglin}, the FS nesting 
which features the structure of $\chi_0$ still plays a non-negligible role in stabilizing magnetic
ordering \cite{YinNMAT}. Indeed, one can see that the nesting vectors connecting the hole FSs 
at $\Gamma$ and the electron FS at $X$ or $Y$ with same orbital characters in Fig.~3b 
roughly coincides with the peak positions of $\chi_m''$ in Fig.~3c. Even though the hole FS is 
absent in Fig.~3e by the electron doping, actually the hole bands are just below $E_F$ as shown
in Fig.~3d so
that the overall nesting condition is not very different from the undoped case. Consequently,
the peak position in $\chi_m''$ plot in Fig.~3f is almost the same as in Fig.~3c, with
only the overall excitation magnitude greatly reduced. The suppressed low energy excitation 
and tendency for a magnetic order rather result from the local two-particle vertex which
includes effects of overall increase of local orbital 
occupations away from the integer filling by doping, which should suppress the fluctuating moment and enhance
charge fluctuations. Zero ordered moment is obtained in the stripe-type AFM calculation for
0.12 $e^-$/Fe doped ML FeSe, in consistence with our $\chi_m''$ result in the PM phase and 
also with the suppressed magnetic order by electron doping found experimentally \cite{YZhou} as mentioned above. 
This large sensitivity of magnetic order on doping therefore results from local correlations, 
which are well described within the DFT+DMFT method.
 
\begin{flushleft}
{\large {\bf Discussion}}
\end{flushleft}
Our result, that strong magnetic order in strained ML FeSe is destroyed by electron doping 
on the level where superconductivity is known to appear, implies the close proximity of magnetism to the 
superconductivity in ML FeSe/SrTiO$_3$, imposing a definite constraint on the electron pairing 
mechanism in this system. Among various pairing scenarios taking into account the absence of
hole FS around $\Gamma$, our results are most consistent with the ``bootstrap" mechanism where
electron FSs and ``incipient" band (hole band below $E_F$) have opposite sign gaps 
($s_{\pm}$) \cite{XChen,Linscheid,DHuang}. This mechanism requires cooperative interplay of attractive
$q \sim (0,0)$ interaction (e.g., by phonon) and repulsive $q \sim (1,0)$ interaction whose existence
is identified in our study as the incommensurate spin excitation. Meanwhile, $q \sim (1,1)$ interaction 
connecting separate electron FSs, as required by other scenarios such as ``nodeless $d$" \cite{Kuroki08,Maier,FWang},
sign-preserving ``s" \cite{Saito,JKang}, and ``bonding-antibonding $s$" \cite{Mazin11,Khodas}, is 
identified from neither spin nor orbital excitations as shown in Fig.~1 and Supplementary Figure 1,
respectively, although non-local
correlations not included in the DFT+DMFT scheme might help stabilize low-energy orbital fluctuations
\cite{Fanfarillo2}. 

Our work casts new light on understanding the dramatic variation of ordered moment
in IBS, including the long standing puzzle of the absence of magnetic order
in bulk FeSe. 
Besides the overall correlation strength as reflected on the size of fluctuating moment,
orbital-specific correlations are also important in determining the magnetic order, as large difference
in intra-orbital charge fluctuation among orbitals, e.g. induced by enhanced crystal-field splitting
in case of bulk FeSe, can give rise to suppressed inter-orbital charge fluctuation and eventually result in
reduced ordered moment of each orbital. 
As our calculated ordered moment of 0.4 $\mu_B$ for bulk
FeSe is still non-zero but certainly smaller than for other materials considered, even tiny amount
of excess electrons generated by intrinsic small excess Fe or Se deficiency \cite{McQueen09prb} might 
easily lead to completely destroyed magnetic order. We expect that other materials which deviate
from the general trend of ordered moment according to the correlation strength and fluctuating
moment as shown in Figure 1 in Ref. 23, such as LiFeAs which also exhibits no magnetic phase, 
can possibly be understood with a similar mechanism. 

In summary, magnetic properties of bulk and tensile-strained ML FeSe are investigated
using DFT+DMFT method. Magnetic susceptibility in the PM state 
indicates suppressed and strongly enhanced magnetic orders
at and near the stripe-type AFM ordering vector for bulk and ML FeSe, respectively.
Bulk FeSe is found to have a pronounced orbital decoupling, i.e., 
strongly reduced inter-orbital charge fluctuations
between $e_g$ and $t_{2g}$ orbitals which result from its large crystal-field splitting
and are manifested by the Hundness of general IBS materials. 
We suggest that the consequently enhanced total charge fluctuation suppresses the static ordering 
of the fluctuating local spin formed by Hund's coupling. On the other hand, magnetic order is strongly  
enhanced in ML FeSe due to enlarged fluctuating moment and slower charge fluctuations caused by more isolated
Fe atoms with the larger lattice constant of the material. We find that the magnetic order disappears after
0.12 $e^-$ doping in ML FeSe along with the hole FSs in the BZ, suggesting a possible relationship
between the magnetic order and the superconductivity in ML FeSe/SrTiO$_3$.

\begin{flushleft}
{\large {\bf Methods}}
\end{flushleft}
{\bf Details of DFT+DMFT calculation.} We perform a systematic analysis for the trend of local correlations 
We use the modern implementation of DFT+DMFT method within all electron embedded
DMFT approach \cite{Haule3}, where in addition to correlated Fe atoms the itinerant states of Se are
included in the Dyson self-consistent equation. The strong correlations on the Fe ion
are treated by DMFT, adding self-energy $\Sigma(\omega)$ on a quasi atomic orbital in real space,
to ensure stationarity of the DFT+DMFT approach. The self-energy $\Sigma(\omega)$ contains all
Feynman diagrams local to the Fe ion. No downfolding or other approximations were used,
and the calculations are all-electron as implemented in Ref. 50, which is based on 
Wien2k \cite{wien2k}.
We employ LDA exchange-correlation functional \cite{LDA1,LDA2}, and the quantum impurity
model was solved by the continuous time quantum Monte Carlo (CTQMC) impurity solver \cite{CTQMC}.
Fixed $U=5.0$ eV and $J=0.8$ eV values are used for all the three materials (see Supplementary Note 2) 
as in the previous work
studying a number of different IBS using the same methodological scheme with the one adopted
in this study \cite{YinNMAT}. These values are also in reasonable agreement with those employed
in a previous LDA+U study for another ferrous material \cite{Baldini}. We use the Slater 
parametrization of the Coulomb interaction in this study, and our $U$ and $J$
parameters are defined with respect to the three Slater parameters in such a way that $F^0= U$, $F^2=112/13~J$,
and $F^4=70/13~J$. Thus this is not to be mistaken for being equivalent to use a single $J$ value averaged 
over different orbitals within more commonly used Kanamori parametrization, and the
anisotropy of interactions among different orbitals is taken into account 
in our calculation with the spherical symmetry assumed \cite{Coulomb}. BZ integration is done on the 14$\times$14$\times$9
k-point mesh for the 2-Fe atom unitcell of bulk FeSe, and equivalent or similar meshes on other structures.
Calculations for PM phases are done at $T=387$ K, and magnetic phases are obtained
at $T=116$ K. All atomic positions are fully optimized with lattice constants fixed to experimental values
\cite{Nomura,McQueen09prb} within DFT+DMFT scheme by minimizing forces
obtained from the derivative of stationary free energy functional as implemented in Ref. 31,
where it is shown how the inclusion of spin fluctuation in DFT+DMFT naturally leads to significantly
better agreement of Se position with experimental values than DFT only calculations.
Optimized atomic positions of As and Se in the internal lattice unit are 0.1537 and 0.2670
with respect to the Fe plane for LaFeAsO and bulk FeSe, which show good agreements with the experimental
values of 0.1517 \cite{Nomura} and 0.2672 \cite{McQueen09prb}. 

\begin{flushleft}
{\large {\bf Data availability}}
\end{flushleft}
The data that support the findings of this study are available from the corresponding
author upon reasonable request.

\begin{flushleft}
{\large {\bf Acknowledgments}}
\end{flushleft}
This research was supported by the Basic Science Research Program through the National Research 
Foundation of Korea (NRF) funded by the Ministry of Science and ICT (2016R1C1B1014715).

\begin{flushleft}
{\large {\bf Author contributions}}
\end{flushleft}
C.-Y.M conceived the project, performed calculations, analyzed data
and wrote the papaer.

\begin{flushleft}
{\large {\bf Competing Interests}}
\end{flushleft}
The author declares no competing interests.

\begin{flushleft}
{\large {\bf References}}
\end{flushleft}


\newpage

\renewcommand{\figurename}{\bf Fig.}
\renewcommand{\tablename}{\bf Table}
\renewcommand{\thetable}{\arabic{table}}
\begin{table}
\caption{{\bf Orbital-resolved charge fluctuations in the PM phase.} 
\normalfont The definition is shown as eq. (1) in the main text. Diagonal 
elements are intra-orbital charge fluctuations which quantify how much
the electron in the orbital is itinerant, while off-diagonal ones are inter-orbital charge
fluctuations which are negative because of the Coulomb repulsion among
orbitals. A number in a parenthesis represents the inter-orbital element between $d_{xz}$
and $d_{yz}$ orbitals, and the number in front of it is the intra-orbital element
of $d_{xz}$ and $d_{yz}$, which are the same. $U=5$ eV and $J=0.8$ eV are used.}

\begin{center}
\begin{tabular}{crrlr}
\hline\hline
            & ~~~$z^2$~~~  & ~~~$x^2-y^2$~ & ~~~~$xz/yz$~~~ &  $xy$~~~~ \\ 
            & \multicolumn{4}{c}{LaFeAsO} \\
\cline{2-5}
 $z^2$      &  0.229       &    0        &  ~~~-0.037     &    -0.003    \\
 $x^2-y^2$  &    0         &    0.212    &  ~~~-0.013     &    {\bf -0.046}    \\
 $xz/yz$    &  -0.037      &   -0.013    & ~~~~0.197 (0.012)&    -0.015    \\
 $xy$       &  -0.003      &{\bf -0.046} &  ~~~-0.015     &     0.222    \\
& \multicolumn{4}{c}{bulk FeSe} \\
\cline{2-5}
 $z^2$      &   0.238      &    0.001    &  ~~~-0.032   &    -0.004    \\
 $x^2-y^2$  &   0.001      &    0.210    &  ~~~-0.014   &   {\bf-0.021}    \\
 $xz/yz$    &  -0.032      &   -0.014    & ~~~~0.184 (0.013)&    -0.014    \\
 $xy$       &  -0.004      &{\bf -0.021} &   ~~~-0.014   &     0.182    \\
& \multicolumn{4}{c}{Monolayer FeSe} \\
\cline{2-5}
 $z^2$      &   0.232      &   -0.002    &  ~~~-0.035   &    -0.005 \\
 $x^2-y^2$  &  -0.002      &    0.208    &  ~~~-0.016   &    -0.023 \\ 
 $xz/yz$    &  -0.035      &   -0.016    & ~~~~0.188 (0.009)&    -0.015  \\
 $xy$       &  -0.005      &   -0.022    &   ~~~-0.015   &     0.182  \\
\hline\hline
\end{tabular}
\end{center}

\label{table I}
\end{table}


\begin{figure}[tp]
\includegraphics[width=0.95\linewidth]{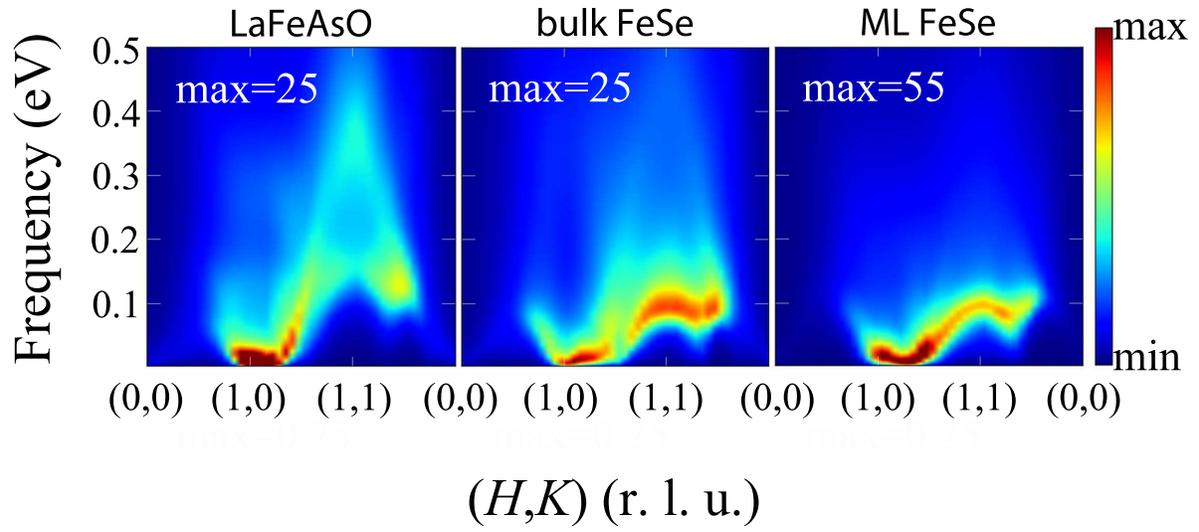}
\caption{{\bf Imaginary part of magnetic susceptibility.} \normalfont For LaFeAsO, bulk FeSe, 
and ML FeSe, respectively. $x$-axis 
is for the momentum transfer ${\bf q}=(H,K,L=1)$ in the reciprocal lattice unit (r. l. u.) of 
one-Fe-unitcell, and $y$-axis is for the frequency.}
\label{fig1}
\end{figure}


\begin{figure}[tp]
\includegraphics[width=0.95\linewidth]{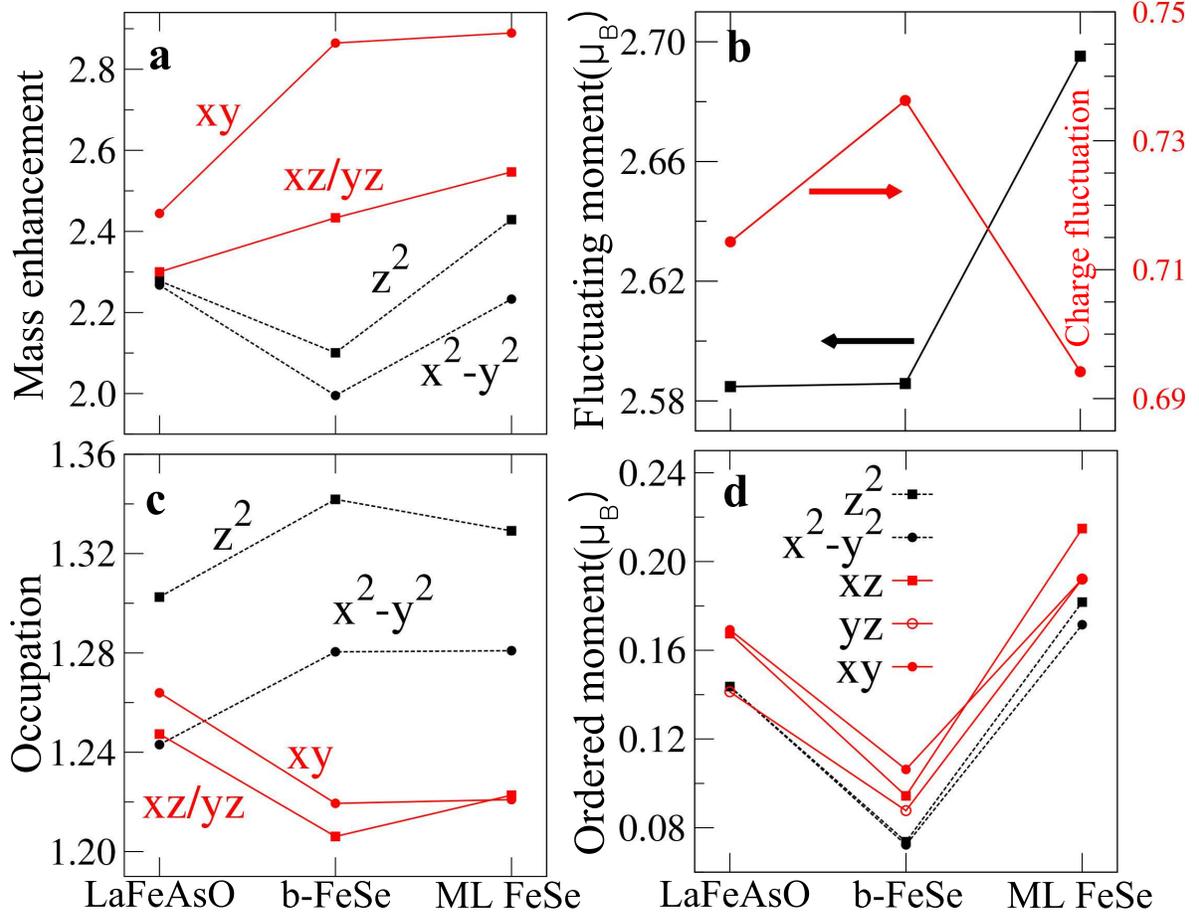}
\caption{{\bf Local quantities.} \normalfont {\bf a}, Orbital-resolved mass enhancement, {\bf b}, fluctuating moment
and local charge fluctuation, {\bf c}, orbital occupations, and {\bf d}, orbital-resolved ordered moments
on an iron atom for LaFeAsO, bulk FeSe (b-FeSe),
and ML FeSe. Ordered moment in {\bf d} are estimated in the AFM phase, while others 
in the PM phase. $U=5$ eV and $J=0.8$ eV are used.}
\label{fig2}
\end{figure}

\newpage

\begin{figure}[tp]
\includegraphics[width=0.85\linewidth]{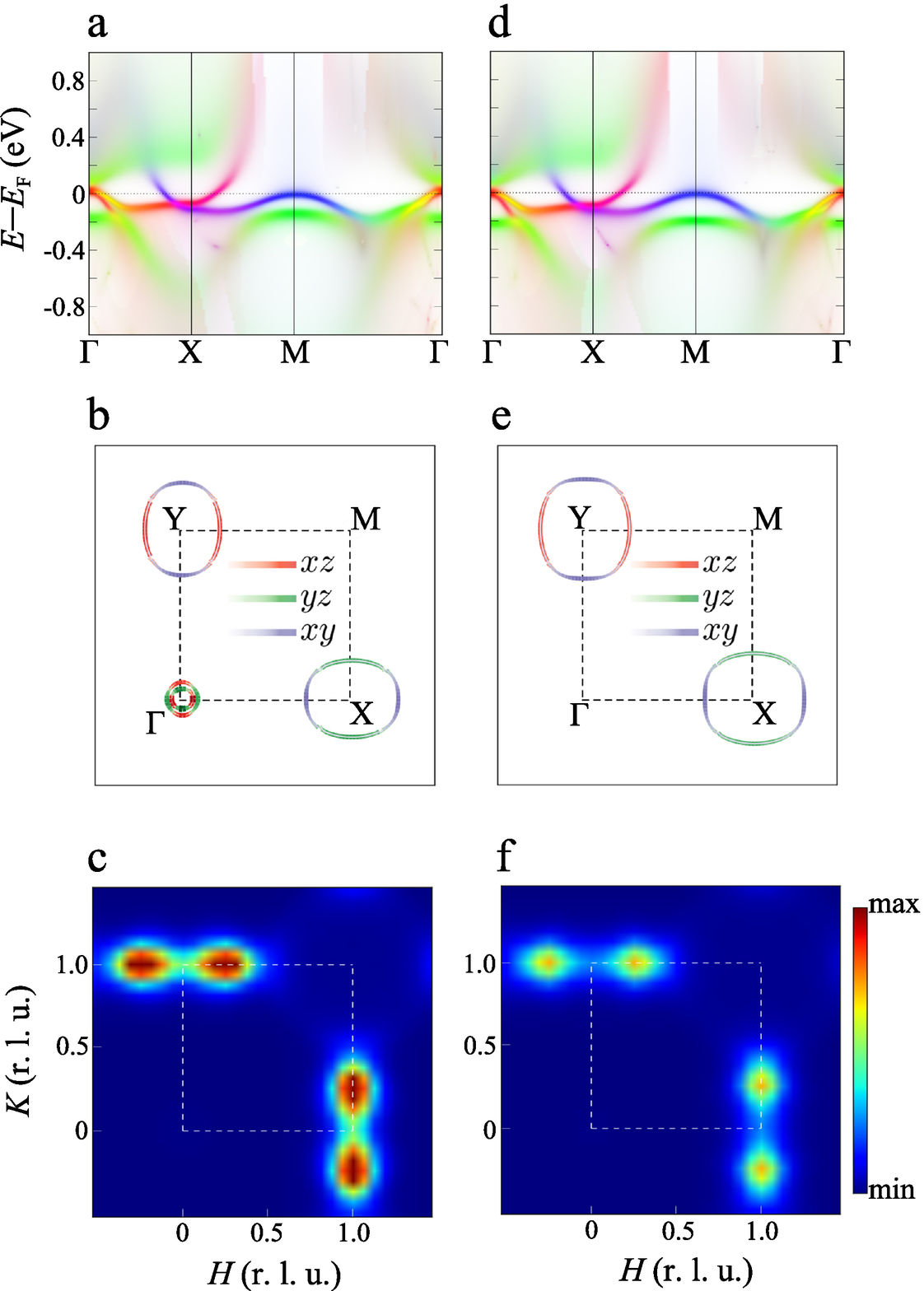}
\caption{{\bf Effect of doping on band structure and spin susceptibility of ML FeSe.}
\normalfont {\bf a}, Orbital-resolved spectral functions $A({\bf k},\omega)$
along the high-symmetry points in the one-Fe-unitcell for the ML FeSe. Red and blue 
represent $d_{xz/yz}$ and $d_{xy}$ components, respectively, while green is for $e_g$ 
orbitals. 
{\bf b}, FS in the two-dimensional BZ of one-Fe-unitcell for ML FeSe, evaluated 
by the real part of the complex energy eigenvalues from the DFT+DMFT calculation. 
Weight of an orbital component is represented by the depth of a color as well as the 
thickness of the line. {\bf c}, $\chi_m''({\bf q},\omega=5$ meV) in the same BZ as that
in {\bf b}. {\bf d}-{\bf f} are counterparts of {\bf a}-{\bf c} for the 0.12 $e^-$/Fe 
doped ML FeSe, respectively.}
\label{fig3}
\end{figure}

\newpage


\begin{thebibliography}{99}
\bibitem{Kanamori} While this relation is valid for the Kanamori parametrization
with the spherical symmetry assumed, actually the Slater parametrization is used 
in this study as described in Methods section and references therein.
\end{thebibliography}
\end{document}